\documentclass{iau}
\usepackage{graphicx}

\title[] 
{Pulsating variables from the OGLE and Araucaria projects}

\author[G. Pietrzy\'nski]   
{G. Pietrzy\'nski$^{1,2}$}

\affiliation{$^1$ Warsaw University Observatory, Al. Ujazdowskie 4, 00-478 Warszawa, Poland\\ 
email: {\tt pietrzyn@astrouw.edu.pl} \\[\affilskip]
$^2$Departamento de Astronom\'ia, Universidad de Concepci\'on, Casilla 160-C, Concepci\'on, Chile
}

\pubyear{2014}
\volume{301}  
\pagerange{1--4}
\jname{Precision Asteroseismology}
\editors{J.A. Guzik, W.J. Chaplin, G. Handler \& A. Pigulski, eds.}
\begin{document}

\maketitle

\begin{abstract}
We present some results of long term studies of pulsating stars conducted in the course of the OGLE
and Araucaria projects. In particular very scarce eclipsing binaries containing pulsating stars are discussed.
Such systems provide a unique opportunity to improve calibration of the cosmic distance scale and
to better calibrate stellar evolutionary models.
\end{abstract}

\firstsection 
\section{OGLE and ARAUCARIA projects}
\underline{Optical Gravitational Lensing Experiment} (OGLE) is one of the biggest astronomical surveys. 
This project has been in operation over the last 21 years. With the current instrumental setup, the 
OGLE team is capable to observe about one billion stars every night. The observations are 
mainly conducted in the Milky Way (bulge and disc) and  Magellanic Clouds. Based on the collected data,
an enormous amount of precise light curves for basically all kinds of  pulsating stars were already 
published: Cepheids (\cite[Soszy\'nski et al.~2008, 2010a]{Sosz08,Sosz10a}), RR Lyrae (\cite[Soszy\'nski et al.~2010b, 2011a]{Szosz10b,Sosz11a}), long period variables (\cite[Soszy\'nski et al.~2011b, 2013]{Sosz11b,Sosz13}), etc., and several new catalogs 
are in preparation. Thanks to an exquisite statistics and high quality of the data, many extremely scarce 
and very interesting objects have been also discovered, including very good candidates for pulsating stars 
in eclipsing binary systems (\cite[Soszy\'nski et al. 2008, 2011a]{Sosz08,Sosz11a}) and a unique sample of eclipsing binary systems composed of clump giants (\cite[Graczyk et al.~2011]{Grac11}).

\underline{Araucaria project.} The main goal of this project is to significantly improve the calibration of the 
cosmic distance scale based on observations of several distance indicators in nearby galaxies. As a first step, 
we performed an optical survey of nine nearby galaxies discovering about 700 Cepheids (\cite[Pietrzy\'nski et al.~2002, 2004, 2006, 2007, 2010a]{Piet02,Piet04,Piet06,Piet07,Piet10a}). 
We also performed infrared 
observations of discovered Cepheids (\cite[Gieren et al.~2005b, 2006, 2009, 2013]{Gier05b,Gier06,Gier09,Gier13}; \cite[Pietrzy\'nski et al.~2006]{Piet06}), and RR Lyrae stars selected from the OGLE catalogs in the LMC, SMC and
Sculptor galaxies (\cite[Pietrzy\'nski et al.~2008, Szewczyk et al.~2008, 2009]{Piet08,Szew08,Szew09}). Recently, a very important part of the Araucaria project is related to detailed studies of 
eclipsing binary systems discovered by the OGLE project, which have very large potential for precise and accurate 
distance determination and also for improving our knowledge of basic physics of pulsating stars.
During my talk I will focus on some results obtained for some of these systems. 

\section{Setting the zero point for P-L relation of pulsating stars}
The distance to the LMC is widely adopted as the zero point in the calibration of the cosmic distance 
scale. Therefore, a precise and accurate distance determination to this galaxy is paramount for astrophysics in general.

Detached eclipsing double-lined spectroscopic binaries offer a
unique opportunity to measure distances directly (e.g. \cite[Kruszewski and Semeniuk 1999]{KrSe99}).
Recently, we measured distance to the LMC with 2\% precision, based on the analysis of eclipsing binary systems 
composed of red clump giants (\cite[Pietrzy\'nski et al.~2013]{Piet13}). For eight such  systems linear sizes of both components 
were measured with 1\% accuracy based on modeling of high-quality photometric light curves obtained by the OGLE project 
and radial velocity curves constructed by the Araucaria team. Having first ever discovered late-type eclipsing binaries
(G type), we used a well calibrated relationship between angular diameter and $V-K$ color 
(\cite[Di Benedetto 2005, Kervella et al.~2004]{DiBe05,Kerv04}) and measured corresponding angular sizes of the components of our systems with an accuracy of 
2\%. As the result we obtained the most accurate and reliable LMC distance, which provides a strong basis for 
the determination of the Hubble constant with an accuracy of about 3\%. At present, we are working 
on improving the surface brightness-color calibration and measuring the LMC distance with an accuracy of 1\%.

Since the OGLE group constructed outstanding period-luminosity (P-L) relations for several different pulsating stars 
in the Large Magellanic Cloud (e.g. \cite[Soszy\'nski et al.~2010a, 2011b, 2013]{Sosz10a,Sosz11b,Sosz13}), our distance determination provides also an opportunity to 
calibrate uniform fiducial P-L relations for basically all P-L relations of pulsating stars used 
for distance determination.

\section{Cepheids in eclipsing binaries}
The OGLE project provided also very good candidates for classical Cepheids in eclipsing systems (e.g. \cite[Soszy\'nski et al.~2008]{Sosz08}).
Such systems provide a unique opportunity to measure precisely and accurately stellar parameters of Cepheids. In consequence, 
they provide very strong constraints on stellar evolutionary and pulsation models.
One can also measure distances to such targets using three independent techniques: P-L relation, Baade-Wesselink 
method, and eclipsing binaries described above. Comparing the independent distances, the potential 
systematic errors associated with each of these methods can be precisely traced out. Because of the 
huge potential of these systems for improving our capability of measuring distances with classical Cepheids and to better understand 
basic physics of these stars, as a part of the Araucaria project 
we started a long-term program to characterize them. In 2010 we confirmed that one of the OGLE candidates --- OGLE-LMC-CEP-227 --- is indeed a physical 
system containing a Cepheid. Based on high-quality data, we measured the dynamical mass of the Cepheid with an 
accuracy of 1\% (\cite[Pietrzy\'nski et al.~2010b]{Piet10b}). Recently, we significantly improved the accuracy of determination of 
the physical parameters of this system and measured directly the p factor for the Cepheid (\cite[Pilecki et al.~2013]{Pile13}).
This analysis complements our previous study on the calibration of the p factor (\cite[Gieren et al.~2005a, Nardetto et al.~2011, Storm et al.~2011]{Gier05a,Nard11,Stor11}). \cite{Piet11} measured the dynamical mass of another 
Cepheid in an eclipsing system with similar accuracy. These results already triggered several theoretical investigations 
(\cite[Cassisi \& Salaris 2011, Neilson et al.~2011, Prada Moroni et al.~2012, Marconi et al.~2013]{CaSa11,Neil11,Prad12,Mark13}).

Unfortunately, there are no  Cepheids in eclipsing binary systems known so far in the Milky Way. However, some of the 
Cepheids in binary systems are sufficiently close to observe them interferometrically (\cite[Gallenne et al.~2013]{Gale13}). Combining spectroscopic 
and interferometric data one should be also able to precisely measure distances, masses and other physical parameters for 
several Cepheids in the Milky Way (\cite[Gallenne et al.~2013]{Gale13}; see also Gallenne et al., these proceedings).

\section{Binary Evolution Pulsating stars}
After many years of intense searching for an RR Lyrae star in an eclipsing binary system 
a very good candidate, OGLE-BLG-RRLYR-02792, was discovered by the OGLE team (\cite[Soszy\'nski et al.~2010b]{Sosz10b}).
Together with high resolution spectra obtained by the Araucaria team, we modeled this system
and determined its physical parameters. Surprisingly, the mass of the primary component, 
pulsating star classified as an RR Lyrae star based on properties of its light curve, turned 
out to be 0.261 $\pm$ 0.015 $M_{\odot}$ (\cite[Pietrzy\'nski et al.~2012]{Piet12}). This is of course incompatible with the theoretical predictions for a 
classical horizontal branch star evolving through the instability strip. However, the relatively short 
orbital period of this system (15.24 days) suggests that mass exchange between the components should have occurred
during its evolution. 

Inspired by this possibility, we calculated a series of models for Algol systems and found that a system which initially contained two stars with masses $M_1 = 1.4\,M_{\odot}$ and $M_2 = 0.8\,M_{\odot}$ orbiting each other with an initial period of 2.9 days would, after 5.4 Gyr of evolution, have exchanged mass between the components as classical Algols do, and today would form a system very similar to RRLYR-02792 (e.g. with $M_1 = 0.268\,M_{\odot}$, $M_2 = 1.665\,M_{\odot}$, and $P_{\rm orb}=15.9$ days). We therefore conclude that the primary component of our observed system is not a classical RR Lyrae star with its well-known internal structure, but rather a star which possesses a partially degenerated helium core and a small hydrogen-rich envelope (shell burning). The primary component has lost most of its envelope to the secondary during the red giant branch phase due to the mass exchange in a binary system, and is now crossing the main instability strip during its evolution
towards the hot subdwarf region of the H-R diagram (\cite[Pietrzy\'nski et al.~2012]{Piet12}). Pulsational properties of this star 
were investigated by \cite{Smol13}. 
As a result, we  discovered a new evolutionary channel of producing binary evolution pulsating (BEP)
stars --- new inhabitants of the classical instability strip. They mimic classical RR Lyrae variables, but have a completely different origin.
Since the primary components can have very different masses during the mass exchange, they can be expected in different 
regions of the instability strip. Very recently, \cite{Maxt13} discovered another BEP star crossing the instability 
strip of $\delta$~Scuti stars. 
Since close binary systems composed of two intermediate-mass stars orbiting each other 
with periods of a few days are relatively frequent, the newly discovered evolutionary channel should produce a significant fraction of the white dwarfs in the Universe. 

\section{Acknowledgements}
We gratefully acknowledge financial support for this work from the Polish       
National Science Center grant MAESTRO DEC-2012/06/A/ST9/00269.

\end{document}